\documentclass[runningheads]{llncs}
\usepackage{graphicx}
\usepackage{xstring}
\graphicspath{ {./images/} }
\usepackage{caption}
\usepackage{subcaption}
\usepackage{amsmath}
\usepackage{tikz}
\usepackage{comment}

\begin{document}
\title{Quantifying Credit Portfolio sensitivity to asset correlations with interpretable generative neural networks}
\titlerunning{Credit Portfolio correlation sensitivity with interpretable generative NN}
%
\author{Sergio Caprioli\inst{1} \and
Emanuele Cagliero\inst{2} \and 
Riccardo Crupi\inst{2}\orcidID{0009--0005--6714--5161}}
\authorrunning{S. Caprioli et al.}
%
\institute{Intesa Sanpaolo S.P.A., Milano MI 20121, Italy
\email{sergio.caprioli@intesasanpaolo.com}\\
\and
Intesa Sanpaolo S.P.A., Torino TO 10121, Italy\\
\email{\{name.surname\}@intesasanpaolo.com}}

\maketitle              
%


\begin{abstract}
In this research, we propose a novel approach for the quantification of credit portfolio Value-at-Risk (VaR) sensitivity to asset correlations with the use of synthetic financial correlation matrices generated with deep learning models. In previous work Generative Adversarial Networks (GANs) were employed to demonstrate the generation of plausible correlation matrices, that capture the essential characteristics observed in empirical correlation matrices estimated on asset returns. Instead of GANs, we employ Variational Autoencoders (VAE) to achieve a more interpretable latent space representation. Through our analysis, we reveal that the VAE latent space can be a useful tool to capture the crucial factors impacting portfolio diversification, particularly in relation to credit portfolio sensitivity to asset correlations changes.

\end{abstract}
\section{Introduction}

\subsection{Credit Portfolio concentration risk} \label{cpm}
One of the most adopted models to measure the credit risk of a loan portfolio was proposed in \cite{vasicek} and it is currently a market standard used by regulators for capital requirements \cite{EU_reg}. This model provides a closed-form expression to measure the risk in the case of asymptotic single risk factor (ASRF) portfolios. The ASRF model is portfolio-invariant, i.e., the capital required for any given loan only depends on the risk of that loan, regardless of the portfolio it is added to. Hence the model ignores the concentration of exposures in bank portfolios, as the idiosyncratic risk is assumed to be fully diversified.
Under the Basel framework, Pillar I capital requirements for credit risk do not cover concentration risk hence banks are expected to autonomously estimate such risk and set aside an appropriate capital buffer within the Pillar II process \cite{eba_guidelines}.

A commonly adopted methodology of measuring concentration risk, in the more general case of a portfolio exposed to multiple systematic factors and highly concentrated on a limited number of loans, is to use a Monte Carlo simulation of the portfolio loss distribution under the assumption reported in \cite{imf_model}. The latter states that the standardized value of the $i$-th counterparty, $V_i$, is driven by a factor belonging to a set of macroeconomic Gaussian factors $\{Y_j\}$ and by an idiosyncratic independent Gaussian process $\varepsilon_i$:
\begin{equation} \label{ASRF}
V_i = \rho_i Y_j + \sqrt{1 - \rho_i^2} \varepsilon_i = \sum_f \rho_i \alpha_{j,f} Z_f + \sqrt{1 - \rho_i^2}\varepsilon_i
\end{equation}
through a coefficient $\rho_i$. The systematic factors $\{Y_j\}$ are generally assumed to be correlated, with correlation matrix $\boldsymbol{\Sigma}$. The third term of Eq. \ref{ASRF} makes use of the spectral decomposition $\boldsymbol{\Sigma} = \boldsymbol{\alpha} \boldsymbol{\alpha}^{T}$ to express $V_i$ as a linear combination of a set of uncorrelated factors $\{Z_f\}$, allowing for a straightforward Monte Carlo simulation.

The bank's portfolio is usually clustered into sub-portfolios that are homogeneous in terms of risk characteristics (i.e. industrial sector, geographical area, rating class or counterparty size). A distribution of losses is simulated for each sub-portfolio and the Value at Risk (VaR) is calculated on the aggregated loss.

The asset correlation matrix $\boldsymbol{\Sigma}$ is a critical parameter for the estimation of the sub-portfolio loss distribution, that is the core component for the estimation of the concentration risk. Therefore it is worth assessing the credit portfolio VaR sensitivity to that parameter.

\subsection{Sampling Realistic Financial Correlation Matrices}
As reported in \cite{evolutions}, \begin{quote}
    
    ``markets in crisis mode are an example of how assets correlate or diversify in times of stress. It is essential to see how markets, asset classes, and factors change their correlation and diversification properties in different market regimes. [\ldots] It is desirable not only to consider real manifestations of market scenarios from history but to simulate new, realistic scenarios systematically. To model the real world, quants turn to synthetic data, building artificially generated data based on so-called market generators." \end{quote}

Marti \cite{gmarti} proposed Generative Adversarial Networks (GANs) to generate plausible financial correlation matrices. The author shows that the synthetic matrices generated with GANs present most of the properties observed on the empirical financial correlation matrices estimated on asset returns.
In line with \cite{gmarti} we generated synthetic asset correlation matrices verifying some ``stylized facts" of financial correlations.

We used a different type of neural network, Variational Autoencoders (VAE), to map historical correlation matrices onto a bidimensional ``latent space", also referred to as the bottleneck of the VAE. After training a VAE on a set of historical asset correlation matrices, we show that it is possible to explain the location of points in the latent space. Furthermore, analyzing the relationship between the VAE bidimensional bottleneck and the VaR values computed by the Credit Portfolio Model using different historical asset correlation matrices, we show that the distribution of the latent variables encodes the main aspects impacting portfolio's diversification as presented in \cite{nguyen2018analysis}.

\section{Sensitivity to the Asset Correlation matrix}

\subsection{Data} \label{data}

The dataset contains $n=206$ correlation matrices of the monthly log-returns of $M=44$ equity indices, calculated on their monthly time series from February 1997 to June 2022, using overlapping rolling windows of size 100 months. Historical time series considered are Total Market (Italy, Europe, US and Emerging Markets) and their related sector indices (Consumer Discretionary, Energy, Basic Materials, Industrials, Consumer Staples, Telecom, Utilities, Technology, Financials, Health Care), the source is Datastream.

\subsection{Variational Autoencoder design} \label{vae_design}

An autoencoder is a neural network composed of a sequence of layers (“encoder” E) that perform a compression of the input into a low-dimensional “latent” vector, followed by another sequence of layers (“decoder” D) that approximately reconstruct the input from the latent vector. The encoder and decoder are trained together to minimize the difference between original input and its reconstructed version.


Variational Autoencoders \cite{VAE_bayes} consider a probabilistic latent space defined as a latent random variable $z$ that generated the observed samples $x$. Hence the ``probabilistic decoder" is given by $p(x|z)$ while the ``probabilistic encoder" is $q(z|x)$. The underlying assumption is that the data are generated from a random process involving an unobserved continuous random variable $z$ and it consists of two steps: (1) a value $z_i$ is generated from some prior distribution $p_\theta^*(z)$, (2) a value $\hat{x_i}$ is generated from some conditional distribution $p_\theta^*(x|z)$. Assuming that the prior $p_\theta^*(z)$ and the likelihood $p_\theta^*(x|z)$ come from parametric families of distributions $p_\theta(z)$ and $p_\theta(x|z)$, and that their PDFs are differentiable almost everywhere w.r.t. both $\theta$ and z, the algorithm proposed by \cite{VAE_bayes} for the estimation of the posterior $p_\theta(z|x)$ introduces an approximation $q_\phi(z|x)$ and minimizes the Kullback-Leibler (KL) divergence of the approximate $q_\phi(z|x)$ from the true posterior $p_\theta(z|x)$.         
Using a multivariate normal as the prior distribution, the loss function is composed of a deterministic component (i.e. the mean squared error MSE) and a probabilistic component (i.e. the Kullback-Leibler divergence from the true posterior):


\begin{equation}
\begin{split}
KL = -\frac{1}{2\overline{n}} \sum_{i=1}^{\overline{n}} \sum_{k=1}^{2}{(1+log({\sigma^i_k}^2) - {\mu^i_k}^2 - {\sigma^i_k}^2)} \\
\text{MSE} = \frac{1}{\overline{n}} \sum_{i=1}^{\overline{n}}{\parallel \mathbf{x}_i - \overline{\mathbf{x}}_i \parallel_2^2} = \frac{1}{\overline{n}} \sum_{i=1}^{\overline{n}}{\parallel \mathbf{x}_i - D(E(\mathbf{x}_i))\parallel_2^2} \\
\text{Loss} = \text{MSE} + \beta \cdot \text{KL}
\end{split}
\end{equation}
where $E$ and $D$ are the encoding and decoding map respectively, $E: \mathbf{x} \in \mathbf{R}^{M \times M} \longrightarrow \boldsymbol{\theta} = \{\mu_1, \mu_2, \sigma_1, \sigma_2\} \in \mathbf{R}^4 $, $D: \mathbf{z} \in \mathbf{R}^{2} \longrightarrow \overline{\mathbf{x}} \in \mathbf{R}^{M \times M} $,   $\mathbf{z} = \boldsymbol{\mu} + \boldsymbol{\sigma} \odot \boldsymbol{\varepsilon}$, $\boldsymbol{\mu} = \{\mu_1, \mu_2\}, \boldsymbol{\sigma} = \{ \sigma_1, \sigma_2\}$, $\boldsymbol{\varepsilon}$ is a bivariate standard Gaussian variable, and $\overline{n} < n$ is the number of samples in the training set.

In this equation, $\mu^i_k$ and $\sigma^i_k$ represent the mean and standard deviation of the $k$-th dimension of the latent space for the sample $\mathbf{x}_i$. The loss function balances the MSE, reflecting the reconstruction quality, with $\beta$ times the KL divergence, enforcing a distribution matching in the $2$-dimensional latent space. The KL divergence can be viewed as a regularizer of the model and $\beta$ as the strength of the regularization.

\begin{figure}
     \centering
        \includegraphics[width=\textwidth]{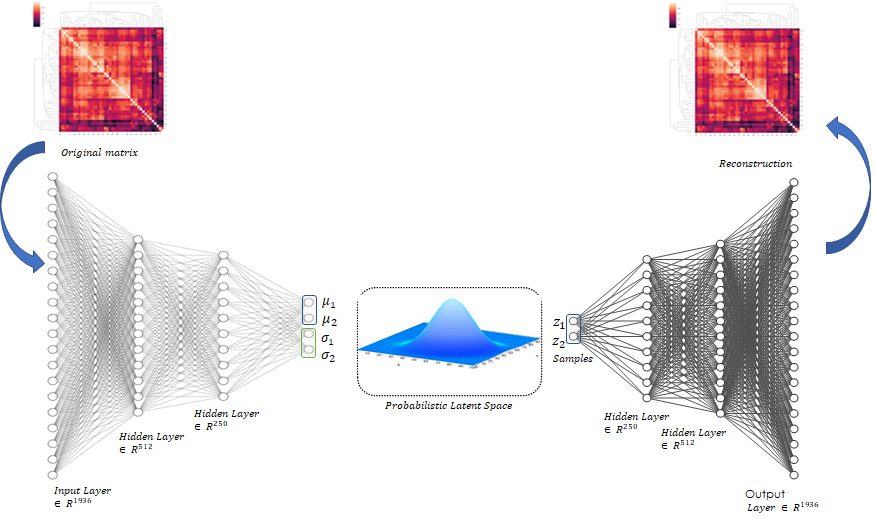}
    
\caption{VAE Framework: The input layer comprises 1936 nodes, corresponding to the $44 \times 44$ matrix input. Subsequently, there are layers with 512, 250, and a central hidden layer with 4 nodes. These values represent the mean and variance of a bivariate gaussian distribution. The decoder receives as input two values sampled from the latent space and it is asked to reconstruct the input. Hence, the architecture is symmetrically mirrored until the output layer, which also has 1936 nodes.}  
 \label{vae_diagram}
\end{figure}

We trained the VAE for 80 epochs using a learning rate of 0.0001 with an Adam optimizer. The structure of the VAE is shown in Fig. \ref{vae_diagram}. We randomly split the dataset described in section \ref{data} in a training sample, used to train the network, and a validation set used to evaluate the performance. We used $30\%$ of the dataset as validation set.



Variational Autoencoders were employed in previous works for financial applications. In particular Brugier and Turinici \cite{vae_bootstrap} proposed a VAE to compute an estimator of the Value at Risk for a financial asset. Bergeron et al. \cite{vae_volatility} used VAE to estimate missing points on partially observed volatility surfaces. Sokol \cite{sokol2022autoencoder} applied VAE for interest rates curves simulation. 

\subsection{Comparison with linear models}

We compared the performances of the Variational Autoencoders with the standard Autoencoder (AE) and with the linear autoencoder (i.e. the autoencoder without activation functions).  

\begin{figure}
     \centering
     \begin{subfigure}[t]{0.4\textwidth}
         \centering
         \includegraphics[width=\textwidth]{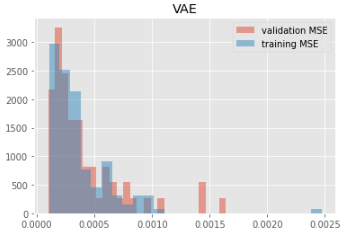}
         \caption{Variational Autoencoder (VAE)}
         \label{vae_loss}
     \end{subfigure}
      \hfill
     \begin{subfigure}[t]{0.4\textwidth}
         \centering
         \includegraphics[width=\textwidth]{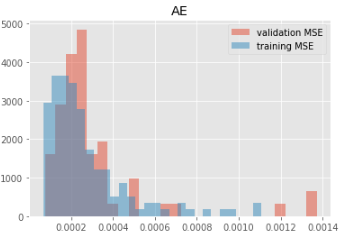}
         \caption{Autoencoder (AE)}
         \label{ae_loss}
     \end{subfigure}
\caption{Histogram of Mean squared error (MSE) of the Autoencoder and Variational Autoencoder on the historical correlation matrices, split into train and validation set.}  

\end{figure}

\begin{figure}
     \centering
     \begin{subfigure}[t]{0.4\textwidth}
         \centering
         \includegraphics[width=\textwidth]{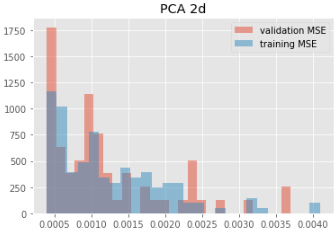}
         \caption{2-dimensional linear autoencoder (PCA 2d)}
         \label{pca2d_loss}
     \end{subfigure}
       \hfill
     \begin{subfigure}[t]{0.4\textwidth}
         \centering
         \includegraphics[width=\textwidth]{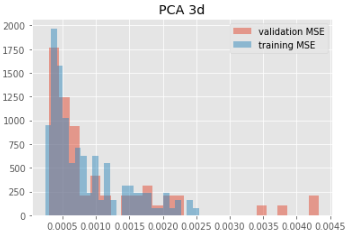}
         \caption{3-dimensional linear autoencoder (PCA 3d)}
         \label{pca3d_loss}
     \end{subfigure}
\caption{Histogram of Mean squared error (MSE) of the linear autoencoders on the historical correlation matrices, split into train and validation set.}  
\end{figure}

The linear autoencoder is equivalent to apply PCA to the input data in the sense that its output is a projection of the data onto the low dimensional principal subspace \cite{pca_ae}.
As shown in Fig. \ref{ae_loss} the autoencoder performs better than VAE (Fig. \ref{vae_loss}), while linear models have lower performances (Fig. \ref{pca2d_loss}) even increasing the dimensions of the latent space (Fig. \ref{pca3d_loss}). Hence, neural networks actually bring an improvement in minimizing the reconstruction error. The generative probabilistic component of VAE decreases the performance when compared to a deterministic autoencoder. On the other hand, it allows to generate new but realistic correlation  matrices in the sense of stylized facts.

\subsection{Latent space interpretability} 

According to Miller \cite{miller2019explanation} and Lipton \cite{lipton2018mythos}:  

\emph{Interpretable} is a model such that an observer can understand the cause of a decision.

\emph{Explanation} is one mode in which an observer may obtain understanding in the latent space, for instance, building a simple surrogate model that mimics the original model to gain a better understanding of the original model's underlying mechanics. 
 
\begin{figure}
     \centering
     \begin{subfigure}[t]{0.49\textwidth}
         \centering
         \includegraphics[width=\textwidth]{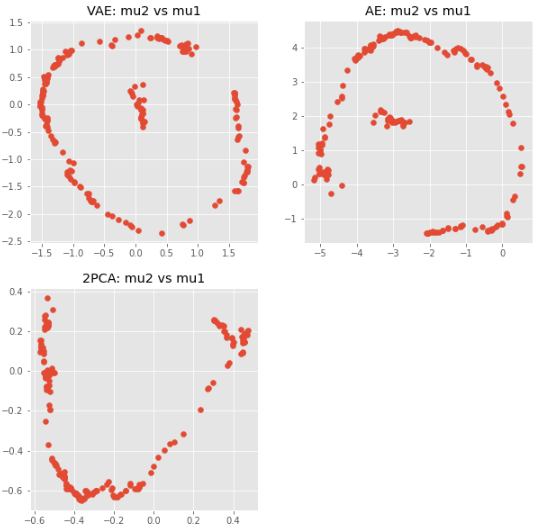}
         \caption{VAE, AE and 2d-PCA latent space}
         \label{scatter_comparison}
     \end{subfigure}
      \hfill
     \begin{subfigure}[t]{0.49\textwidth}
         \centering
         \includegraphics[width=\textwidth]{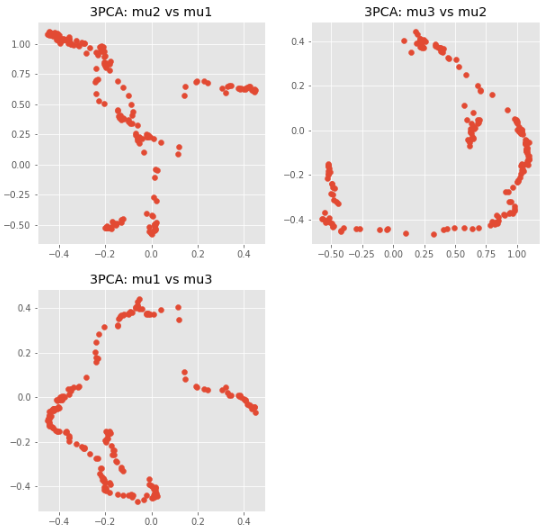}
         \caption{3-d PCA latent space}
         \label{PCA3_scatter}
     \end{subfigure}
\caption{Comparison of the latent space generated with different models. The latent space generated by the VAE and AE are similar while the cluster of points in the middle is recovered only by the 3-dimensional linear autoencoder.}  

\end{figure}

For the sake of our analysis, we refer to the ``interpretability" of the VAE as the possibility to understand the reason underlying the responses produced by the algorithm in the latent space. 
The Variational Autoencoder projected the 206 historical correlation matrices on a two dimensional probabilistic latent space represented by a bivariate normal distribution. As shown in Fig. \ref{scatter_comparison}, the latent space generated by the VAE and AE are similar while the cluster of points in the middle is recovered only by the 3-dimensional linear autoencoder (Fig. \ref{PCA3_scatter}). 

\begin{figure}
  \centering
         \includegraphics[width=0.6\textwidth]{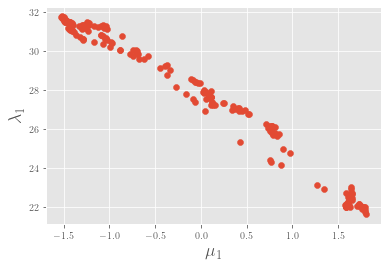}
         \caption{Scatterplot of the first eigenvalue $\lambda_1$ versus the first component of the latent space $\mu_1$, showing a clear negative correlation.}
         \label{first_eig_mu1}

\end{figure}

 In order to understand the rationales underlying such representation, we analysed the relationship of the  encoded values of the original correlation matrices with respect to their eigenvectors $\{\nu_i \mid i=1:M\}$ and eigenvalues $\{\lambda_i \mid i=1:M\}$. It turned out that the first component of the latent space ($\mu_1$) is strongly negatively correlated to the first eigenvalue (Fig. \ref{first_eig_mu1}).
 
As pointed out in \cite{mst_pearson} \begin{quote}“the largest eigenvalue of the correlation matrix is a measure of the intensity of the correlation present in the matrix, and in matrices inferred from financial returns tends to be significantly larger than the second largest. Generally, this largest eigenvalue is larger during times of stress and smaller during times of calm.” \end{quote} Hence, the first dimension of the latent space seems to capture the information related to the rank of the matrix i.e. to the ``diversification opportunities" on the market. 
The interpretation of the second dimension ($\mu_2$) of the latent space turned out to be related to the eigenvectors of the correlation matrix. 
In order to understand the other dimension we consider the cosine similarity $\alpha_t^i$ between the $i$-th eigenvector at time $t$ and its average over time. Formally:
\begin{equation}
    \alpha_{i,t} = \frac{\frac{1}{n}(\sum_{t=1}^n \nu_{i,t})^T  \cdot \nu_{i,t}}{\parallel \frac{1}{n} (\sum_{t=1}^n \nu_{i,t})^T \parallel \parallel  \nu_{i,t} \parallel}
\end{equation}
where $i$ is the number of the eigenvector and $t$ the index of the matrix in the dataset.

\begin{figure}
     \centering
     \begin{subfigure}[b]{0.45\textwidth}
         \centering
         \includegraphics[width=\textwidth]{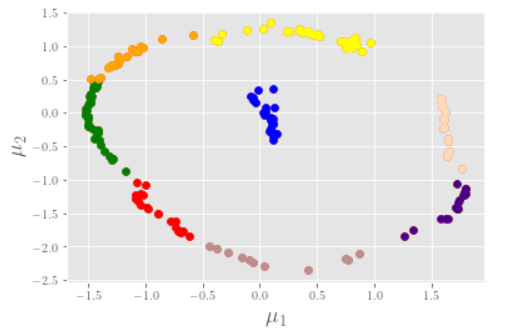}
         \caption{The points on the latent space $(\mu_1, \mu_2)$ representing the historical correlation matrices. The latent space was conventionally partitioned in nine subgroups of data points identified by different colors.}
         \label{scatter_explanation}
     \end{subfigure}
     \hspace{1cm}
     \vspace{1cm}
     \begin{subfigure}[b]{0.45\textwidth}
         \centering
         \includegraphics[width=\textwidth]{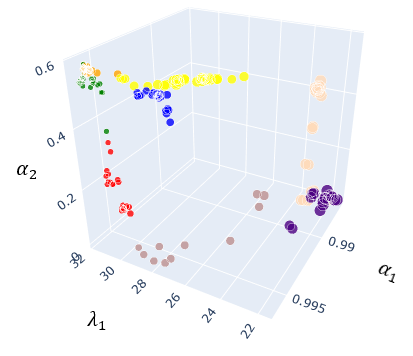}
         \caption{
         The data points of Fig. (a) represented in the parameter space defined by $\alpha_1$, $\alpha_2$, and $\lambda_1$ (also, the size of each dot corresponds to the value of $\lambda_2$). The proximity of these data points consistently mirrors the subgroups illustrated in Fig. (a), with matching color schemes. There is a noticeable separation between different subgroups, and this separation is well-defined in most cases.}
         \label{3d_scatter_explanation}
     \end{subfigure}
     \vspace{1cm}
     \begin{subfigure}[b]{0.45\textwidth}
         \centering
         \includegraphics[width=\textwidth]{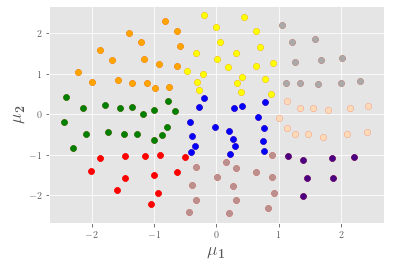}
         \caption{Sampling in the latent space $(\mu_1, \mu_2)$: each point can be decoded into a synthetic correlation matrix. The latent space was conventionally partitioned in nine regions identified by different colors, with the same convention as Fig (a).}
         \label{synt_scatter_explanation}
     \end{subfigure}
     \hspace{1cm}
     \begin{subfigure}[b]{0.45\textwidth}
         \centering
         \includegraphics[width=\textwidth]{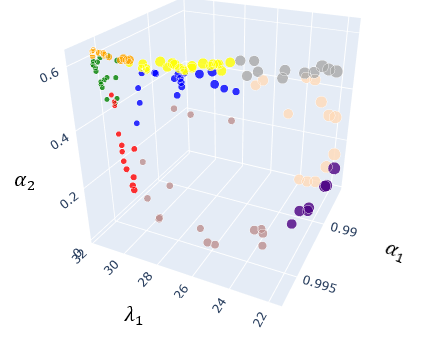}
         \caption{The sampled points of Fig. (c), plotted with the same color schemes in the space formed by $\alpha_1$, $\alpha_2$ and $\lambda_1$  (also, the size of each dot corresponds to the value of $\lambda_2$). Similar observations and considerations can be drawn here as those made for Fig. (a) and Fig. (b).}
         \label{3d_scatter_explanation_synt}
     \end{subfigure}
\caption{The distribution of the distance of the first two eigenvectors from their respective historical average and the distribution of the first eigenvalue characterize the regions of the latent space.}
\label{explanation}

\end{figure}

Let us define $\alpha_1 = \{\alpha_{1,t}\}_{t=1, \ldots n}$, $\alpha_2 = \{\alpha_{2,t}\}_{t=2, \ldots n}$. The data point subgroups observed in the space $(\alpha_1, \alpha_2, \lambda_1)$ can be traced to corresponding subgroups in the latent space $(\mu_1, \mu_2)$, as shown in Fig. \ref{explanation}.


As pointed out in \cite{nguyen2018analysis}, each eigenvector can be viewed as a portfolio weights of stocks that defines a new index which is uncorrelated with the other eigenvectors.
It follows that a change in eigenvectors can impact portfolio diversification. We can conclude that the VAE latent space effectively captures, in two dimensions, the main factors impacting the financial correlations, which is determinant for portfolio diversification.

\subsection{Generating synthetic correlation matrices}

As explained in section \ref{vae_design}, the probabilistic decoder of the VAE allows to generate a ``plausible" correlation matrix starting from any point of the latent space. Hence, we defined a grid of 132 points of the latent space that cover approximately homogeneously an area centered around the origin and including the historical points (Fig. \ref{grid}). For each point on the grid, we used the decoder (i.e. a neural network) to compute the corresponding correlation matrix. Along the lines of \cite{gmarti}, we check whether the following stylized facts of financial correlation matrices hold for both the historical and the synthetic matrices.

\begin{itemize}
  \item{The distribution of pairwise correlations is significantly shifted towards positive values.}
  \item{Eigenvalues follow the Marchenko–Pastur distribution, except for a very large first eigenvalue and a couple of other large eigenvalues.}
  \item{The Perron-Frobenius property holds true (first eigenvector has positive entries and has multiplicity one).}
  \item{Correlations have a hierarchical structure.}
  \item{The Minimum Spanning Tree (MST) obtained from a correlation matrix satisfies the scale-free property.}

\end{itemize}

As shown in Fig. \ref{prop1}, the distributions of pairwise correlations are shifted to the positive both for synthetic and historical matrices. It is worth noting that the synthetic correlations show a more symmetric distribution.
The distributions of the eigenvalues (each averaged respectively over the historical and synthetic matrices) are very similar to each other (Fig. \ref{marchenko}) and can be approximated by a Marchenko-Pastur distribution but for a first large eigenvalue and a couple of other eigenvalues. It is worth noting that the correlation matrices analyzed for our purposes are calculated starting from 44 equity indices (as explained in Section \ref{data}) instead of single stocks as shown in \cite{gmarti}, hence a higher degree of concentration is expected.   
As shown in Fig. \ref{perron_frob} the eigenvector corresponding to the largest (real) eigenvalue has strictly positive components both on historical and synthetic matrices.
In Fig. \ref{hierarchy} we show the heatmap and the dendrogram of two randomly chosen correlation matrices. A hierarchical structure of the correlations can be observed even if, as already pointed out, we are analysing indices instead of single stocks. 
In Fig. \ref{mst} we show the distribution of the degrees of the Minimum Spanning Tree\footnote{For the construction of the MST we followed \cite{mst_pearson}.} calculated on the mean of the historical and synthetic matrices. Both show very few nodes with high degrees while most nodes have degree 1.

\subsection{Quantifying the sensitivity to asset correlations}

\begin{figure}
  \centering
         \includegraphics[width=0.6\textwidth]{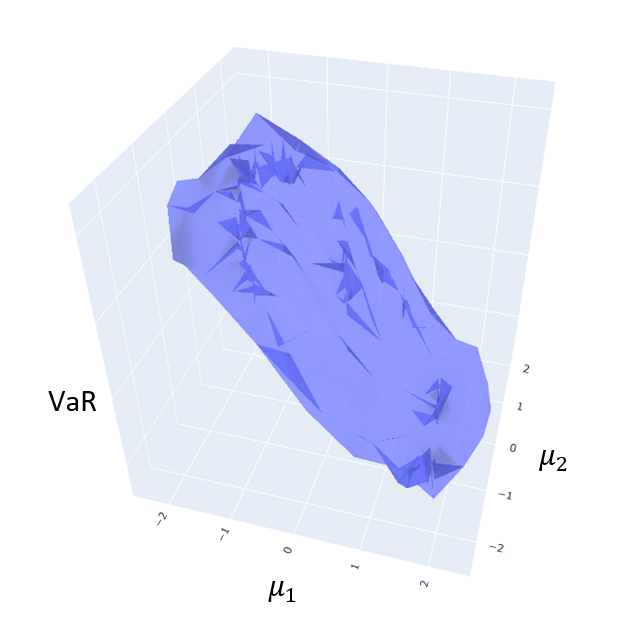}
         \caption{The surface generated from Value at Risk with respect to the points of the 2-dimensional latent space.}
         \label{var_surface}

\end{figure}

For each matrix generated with the VAE probabilistic decoder, we estimated the corresponding VaR according to the multi-factor Vasicek model described in section \ref{cpm}. We used the VaR metric to show a proof of concept of the methodology and to be aligned to the Economic Capital requirements, but the same rationale can be followed adopting a different risk metric. 
As mentioned in section \ref{cpm}, Vasicek multi-factor model cannot be solved in closed form solution, hence it is necessary to run a Monte Carlo simulation for each generated matrix. We used a stratified sampling simulation with 1 million runs. In each estimation, the parameters of the model and portfolio exposures are held constant. Running the simulation for every sampled point of the latent space, we derived the VaR surface of Fig. \ref{var_surface}.

\begin{figure}
  \centering
         \includegraphics[width=1\textwidth]{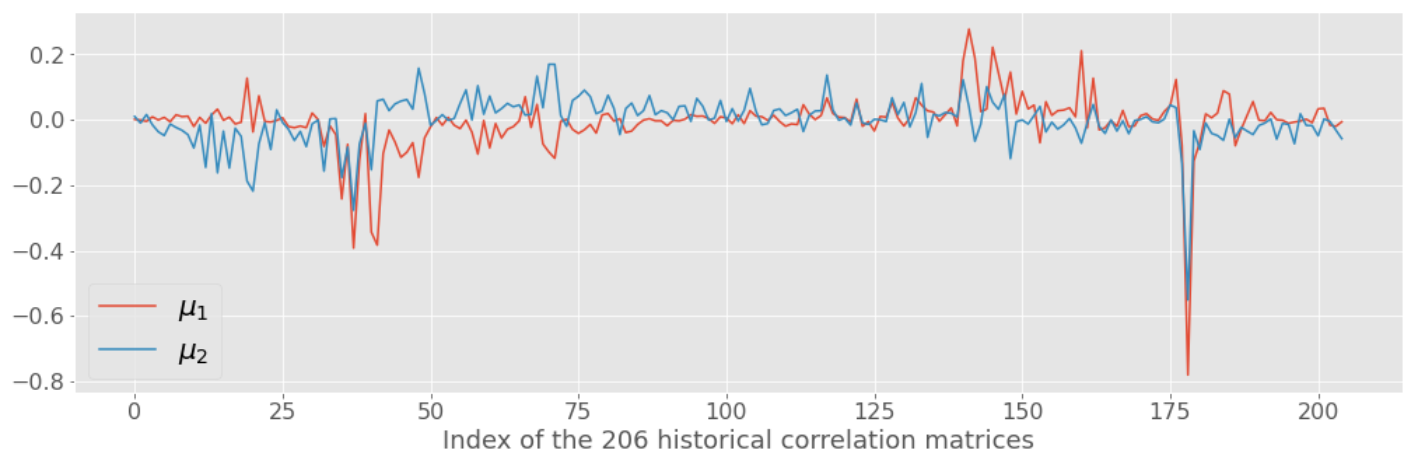}
         \caption{The time series of $\mu_1$ and $\mu_2$, projections of the 206 historical correlation matrices in the 2-dimensional latent space.}
         \label{mu_1_2_time_series}

\end{figure}

\begin{figure}
  \centering
         \includegraphics[width=0.75\textwidth]{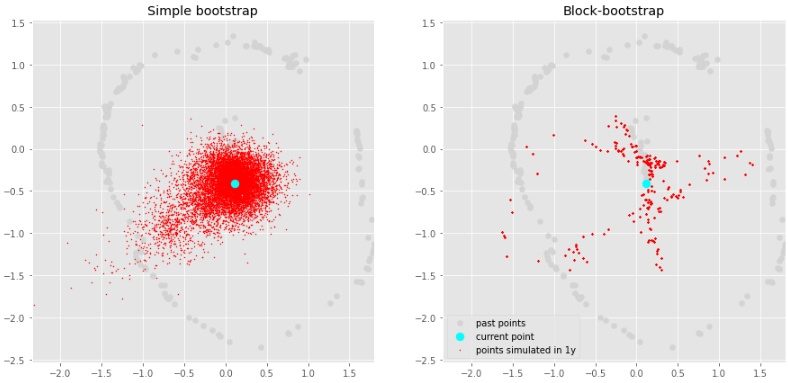}
         \caption{Using simple bootstrap (on the left) or block-bootstrap (with 11 monthly steps) on the ``compressed" representation of the correlation matrices, we estimated the distribution of the possible variation of the current matrix on 1-year time horizon.}
         \label{bootstrap}

\end{figure}

To obtain an estimate of the sensitivity of the VaR to future possible evolutions of the correlation matrix, we ``bootstrapped'' (see Fig. \ref{bootstrap}) the historical time series of the points in the 2-dimensional latent space. We used a simple bootstrapping \cite{abney2002bootstrapping} and a block-bootstrapping technique \cite{mader2013block} on the differences' time series of the two components of the VAE latent space, $\mu_1$ and $\mu_2$  (depicted in Fig. \ref{mu_1_2_time_series}).

Interpolating the estimated VaR over the randomly sample grid (Fig. \ref{var_surface}) we can derive the Value at Risk corresponding to any point of the latent space. Hence, for each point belonging to the distribution of correlations changes over a 1-year time horizon estimated via bootstrap, we can calculate the corresponding VaR without recurring to the Monte Carlo simulation. 

In this way, we obtained the VaR distribution related to the possible variations of correlation matrices on a defined time horizon.  

\section{Conclusions}

In this work we applied a Variational Autoencoder for generating realistic financial matrices that we used as input for the estimation of credit portfolio concentration risk estimated with a multi-factor Vasicek model. We deviated from the methodology proposed by G. Marti 2020 \cite{gmarti} who adopted a Generative Adversarial Network, in order to obtain an interpretable model by leveraging the dimensionality reduction provided by VAE. Using as a proof of concept a VAE trained on a dataset composed of 206 correlation matrices calculated on the time series of 44 equity indices using a rolling window of 100 months, we showed how it is possible, even using a small data sample, to derive an interpretation of the latent space that seems aligned to the main aspects driving portfolio diversification \cite{nguyen2018analysis}. 

We exploited the generative capabilities of the VAE to extend the scope of the model beyond the necessarily limited size of the historical sample, generating a larger set of correlation matrices that retain all the realistic features observed in the market, as determined by appropriate tests. We computed the augmented sample of synthetic correlation matrices on a grid in the 2-dimensional VAE latent space and for each synthetic matrix the corresponding credit portfolio loss distribution (and its VaR at a certain percentile) obtained via Monte Carlo simulation under a multi-factor Vasicek model. This way we estimated a VaR surface over the VAE latent space.

Analyzing the time series of the encoded version of the correlation matrices (i.e. the two components of the probabilistic latent space) we easily estimated via bootstrapping, the possible variation of the correlation matrices over a 1-year time horizon. Finally, using the interpolated VaR surface, we were able to estimate the corresponding VaR distribution obtaining a quantification of the impact of the correlations movements on the credit portfolio concentration risk.  
This approach gives rapid estimation of risk without depending on the extensive computations of Monte Carlo simulation, and it does so in a compressed, easy-to-visualize space that captures several aspects of market dynamics.
Our analysis provides clear indications that the capabilities of realistic data-augmentation provided by Variational Autoencoders combined with the ability to obtain model interpretability can prove useful for risk management purposes, when addressing the sensitivity of models on a structured multidimensional market data as the correlation matrix.

\section*{Disclaimer\label{sec:disclaimer}}
The views and opinions expressed within this paper are those of the authors and do not necessarily reflect the official policy or position of Intesa Sanpaolo. Assumptions made in the analysis, assessments, methodologies, models and results are not reflective of the position of any entity other than the authors.

\nocite{*}
\bibliography{bib/ref.bib}
\bibliographystyle{acm}

\newpage

\appendix
\section{Additional figures}
\begin{figure}
  \centering
         \includegraphics[width=0.4\textwidth]{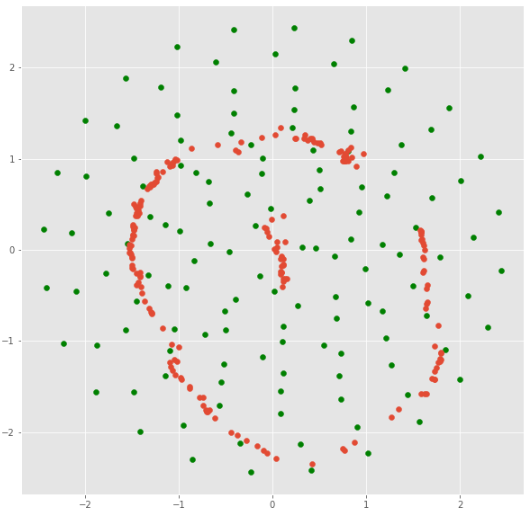}
         \caption{A grid of randomly chosen points (green) on the VAE bottleneck (historical points are in red). }
         \label{grid}

\end{figure}

\begin{figure}
     \centering
     \begin{subfigure}[b]{0.45\textwidth}
         \includegraphics[width=0.75\textwidth]{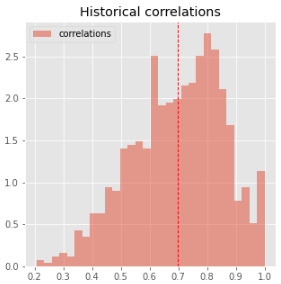}
         \caption{Historical matrices: pairwise correlations.}
         \label{prop1_hist}
     \end{subfigure}
      \hfill
      \begin{subfigure}[b]{0.45\textwidth}
         \includegraphics[width=0.75\textwidth]{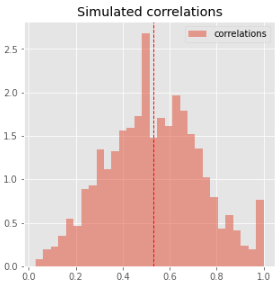}
         \caption{Synthetic matrices: pairwise correlations.}
         \label{prop1_sim}
     \end{subfigure}
     \caption{Distribution of pairwise correlations is significantly shifted to the positive}
\label{prop1}
\end{figure}

\begin{figure}
     \centering
     \begin{subfigure}[b]{0.45\textwidth}
         \includegraphics[width = 1\textwidth]{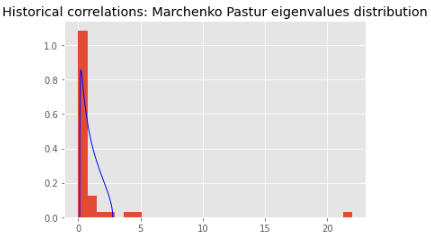}
         \caption{Historical matrices: Marchenko-Pastur eigenvalues distribution.}
         \label{marchenko_hist}
     \end{subfigure}
      \hfill
      \begin{subfigure}[b]{0.45\textwidth}
         \includegraphics[width = 1\textwidth]{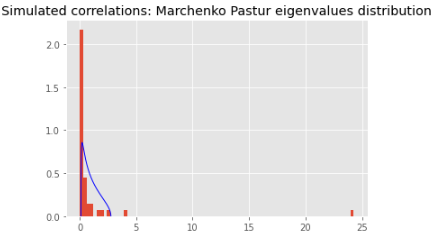}
         \caption{Synthetic matrices: Marchenko-Pastur eigenvalues distribution.}
         \label{marchenko_sim}
     \end{subfigure}
     \caption{Eigenvalues are compatible with the Marchenko–Pastur distribution (that describes the spectrum of a random matrix), with the exception of a very large first eigenvalue and a couple of other large eigenvalues.}
\label{marchenko}
\end{figure}

\begin{figure}
     \centering
     \begin{subfigure}[b]{0.45\textwidth}
         \includegraphics[width = 1\textwidth]{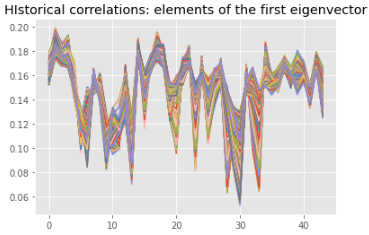}
         \caption{Historical matrices: Perron-Frobenius property.}
         \label{perron_hist}
     \end{subfigure}
      \hfill
      \begin{subfigure}[b]{0.45\textwidth}
         \includegraphics[width = 1\textwidth]{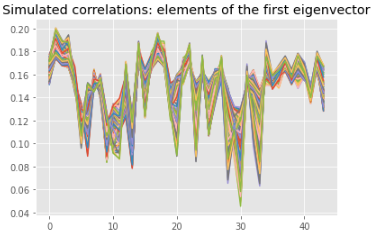}
         \caption{Synthetic matrices: Perron-Frobenius property. }
         \label{perron_sim}
     \end{subfigure}
     \caption{Perron-Frobenius property (first eigenvector has positive entries).}
\label{perron_frob}
\end{figure}

\begin{figure}
     \centering
     \begin{subfigure}[b]{0.45\textwidth}
         \includegraphics[width = 1\textwidth]{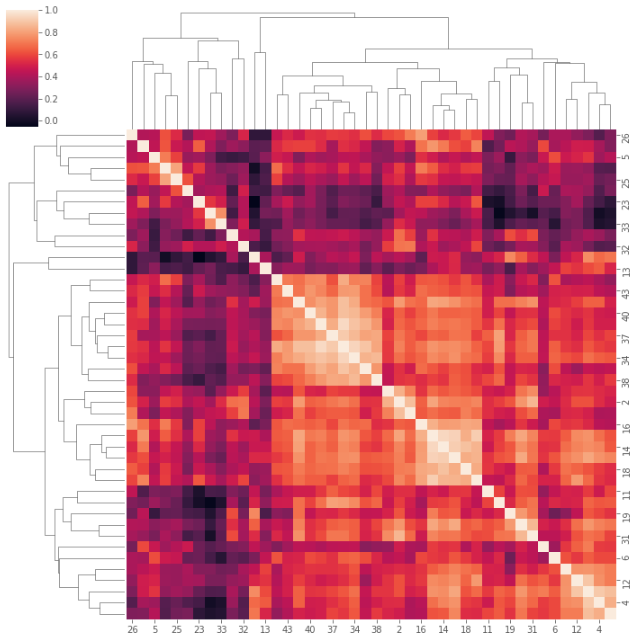}
         \caption{A randomly chosen historical matrix: heatmap and dendrogram.}
         \label{hierarchy_hist}
     \end{subfigure}
      \hfill
      \begin{subfigure}[b]{0.45\textwidth}
         \includegraphics[width = 1\textwidth]{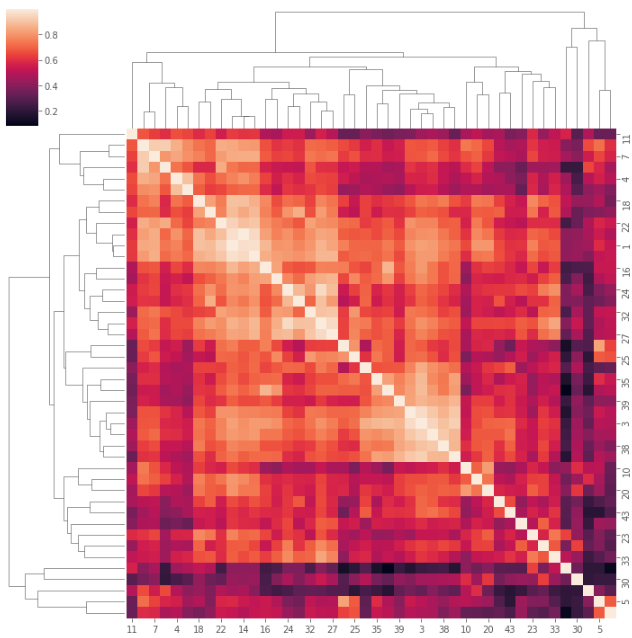}
         \caption{A randomly chosen synthetic matrix: heatmap and dendrogram. }
         \label{hierarchy_sim}
     \end{subfigure}
     \caption{Hierarchical structure of correlations.}
\label{hierarchy}
\end{figure}

\begin{figure}
     \centering
     \begin{subfigure}[b]{0.45\textwidth}
         \includegraphics[width = 1\textwidth]{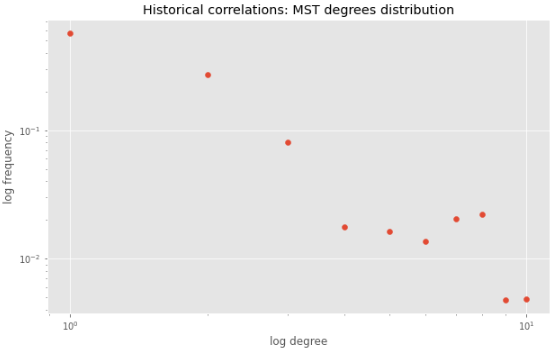}
         \caption{Historical correlations: distribution of MST degrees.}
         \label{mst_hist}
     \end{subfigure}
      \hfill
      \begin{subfigure}[b]{0.45\textwidth}
         \includegraphics[width = 1\textwidth]{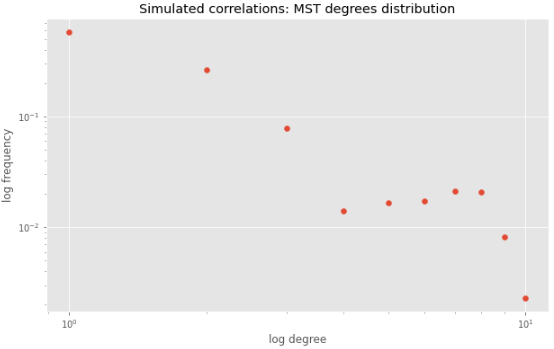}
         \caption{Synthetic correlations: distribution of MST degrees.}
         \label{mst_sim}
     \end{subfigure}
     \caption{Scale-free property of the corresponding MST.}
\label{mst}
\end{figure}

\end{document}